\documentclass[prl,twocolumn,showpacs,amsmath,amssymb]{revtex4}
\usepackage{graphicx}
\usepackage{dcolumn}
\usepackage{epsfig}

\newcommand{\be}{\begin{equation}}
\newcommand{\ee}{\end{equation}}
\newcommand{\bea}{\begin{eqnarray}}
\newcommand{\eea}{\end{eqnarray}}
\newcommand{\nn}{\nonumber}
\newcommand{\la}{\lambda}

\def\Tr{{\mbox{Tr}}}
\def\Pf{{\mbox{Pf}}}
\def\re{{\Re e}}

\begin{document}

\title{A Wigner Surmise for Hermitian and Non-Hermitian Chiral Random Matrices
}
\author {G. Akemann$\,{}^\sharp$, E. Bittner$\,{}^\flat$,
M.J.~Phillips$\,{}^\sharp$, and L.~Shifrin$\,{}^\sharp$
}

\affiliation{${}^{\sharp}$Department of Mathematical Sciences
\& BURSt Research Centre, Brunel University West London,
Uxbridge UB8 3PH, United Kingdom\\
       ${}^\flat$Institute for Theoretical Physics and Centre for Theoretical
Sciences (NTZ), University Leipzig,
        P.O. Box 100 920, D-04009 Leipzig, Germany}

\date   {\today}

\begin{abstract}
We use the idea of a Wigner surmise to compute approximate distributions of
the first 
eigenvalue in chiral Random Matrix Theory, for both real and complex
eigenvalues. Testing against known results
for zero and maximal non-Hermiticity in the microscopic large-$N$ limit
we find an excellent agreement, valid for a small number of
exact zero-eigenvalues. New compact expressions are derived for real
eigenvalues in the orthogonal and symplectic classes, and at
intermediate non-Hermiticity for the unitary and symplectic classes.
Such individual Dirac eigenvalue
distributions are a useful tool in Lattice Gauge Theory and
we illustrate this by showing that our new
results can describe 
data from two-colour QCD simulations with chemical potential
in the symplectic class.

\end{abstract}



\maketitle

1. {\it Motivation.}
Probably
one of the most used predictions of Random Matrix Theory (RMT) is the
so-called Wigner surmise (WS) describing the universal repulsion of energy
levels in many systems in nature, including neutron scattering, quantum
billiards and elastomechanical modes in crystals \cite{GMW}.
For large matrices, the nearest-neighbour (nn) spacing  distribution
$p^{(\beta)}(s)$ is universal and 
only depends on the repulsion strength which takes discrete values
$\beta\!=\!1,2,4$ 
for the three classical Wigner-Dyson (WD) ensembles.
It can be computed with surprising accuracy using $2\times2$ matrices, which
is the WS. Although simple arguments discussed in \cite{Mehta} lead to this
rule for $\beta\!=\!1$, such an approximation is by no means obvious.

The extension from WD 
to non-Hermitian RMT introduced long ago by Ginibre \cite{Gin}
has become a very active field in the past decade, in particular due to
applications in open quantum systems, see
\cite{reviewFS} for references and other applications.
Here the spacing is known only for the class with broken time-reversal
($\beta=2$) and has been applied in Lattice Gauge Theory (LGT) \cite{MPW}.
However, a simple surmise based on $2\times 2$ matrices does not work here.

In this paper we investigate the existence of a surmise for
the smallest eigenvalue in
chiral RMT and its non-Hermitian extensions. These have become
relevant due to applications in Quantum Chromodynamics
(QCD) initiated by \cite{SV93} and extended to non-Hermitian QCD at finite
quark chemical potential $\mu$
\cite{Steph}. QCD at strong coupling is a notoriously difficult theory,
and the chiral RMT approach has become an important tool for
LGT with exact chiral fermions \cite{Edwards,Bloch}.
For non-Hermitian QCD the complex action
hampers a straightforward LGT approach, see
\cite{QCDmu} for a recent discussion and references.
Here RMT predictions remain possible for various
quantities \cite{James,AOSV,KimJac}.

In this paper we will show that an excellent approximation for the 1st
non-zero eigenvalue is possible using a simple  $2\times(2+\nu)$ matrix
calculation, capturing the repulsion of a small number $\nu$ of zero
eigenvalues. Being localised and non-oscillatory the 1st eigenvalue is much
more suitable for LGT than the spectral density, compare e.g. \cite{Bloch} and
\cite{ABSW}.
Our surmise fills some gaps in predictions for real eigenvalues in the
orthogonal and symplectic classes ($\beta=1,4$) \cite{DN98},
where until very recently numerically generated RMT had to be used for
comparison \cite{ITEP}. We also provide new
predictions for intermediate non-Hermiticity and test them against QCD-like LGT
data from \cite{AB}. This further completes the picture, compared to previous
approximations \cite{ABSW} ($\beta=2$) based on a Fredholm determinant 
expansion \cite{AD03}, and exact results at maximal non-Hermiticity
\cite{APS} ($\beta=2,4$).


2. {\it Level spacing in the WD class.}
We recall here the success of a WS for Hermitian, and
its failure for non-Hermitian, WD ensembles.
The WD partition function for an $N\times N$ Hermitian matrix $H$
with real, complex or
quaternion real entries is given terms of eigenvalues by
\be
{\cal Z}_{W\!D}^{(\beta)}=\int dH \mbox{e}^{-\Tr HH^\dag}
\sim\int_{\mathbb{R}}
\prod_{j=1}^N d\la_j
\mbox{e}^{-\la_j^2}|\Delta_N(\la)|^\beta.
\label{ZWD}
\ee
The Jacobians of the corresponding ensembles which are called GOE, GUE and GSE 
($\beta\!=\!1,2$ and $4$) include the
Vandermonde determinant, $\Delta_N(\la)\equiv\prod_{k>l}^N(\la_k-\la_l)$.

The large-$N$ nn spacing in the bulk of the spectrum
can be computed approximately from $N\!=\!2$ (WS) by
inserting $\delta(|\la_1-\la_2|-s)$
in ${\cal Z}_{W\!D}^{(\beta)}$:
\be
p_{W\!S}^{(\beta)}(s)=a_\beta\, s^\beta \exp[-b_\beta\, s^2] \ .
\label{WS}
\ee
The constants $a_\beta, b_\beta$  follow from fixing
the norm and first moment to unity (see e.g. in \cite{GMW}).
The latter can always be achieved from $\int_0^\infty ds\,s\,\hat{p}(s)=m$ by
rescaling $p^{(\beta)}(s)=m\hat{p}^{(\beta)}(ms)$. This
fixes the scale compared with $N=\infty$.

\begin{figure}[t]
\centerline{
  \epsfig{figure=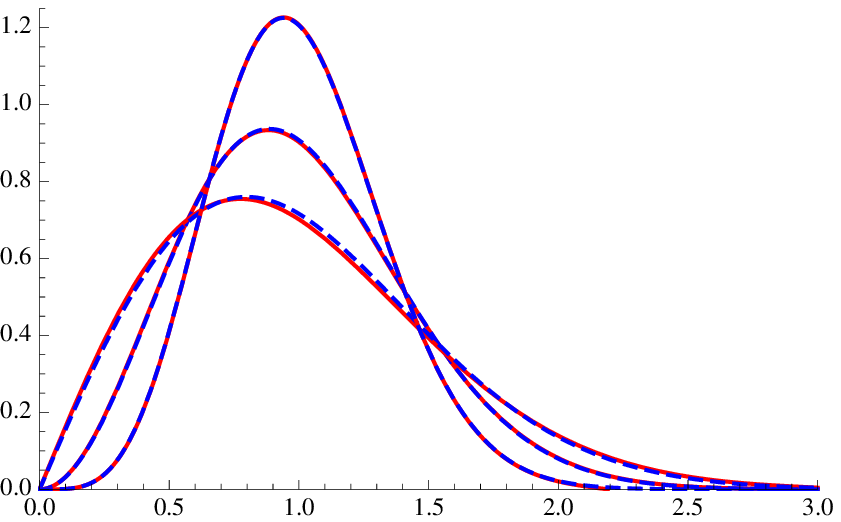,height=28mm}\hspace*{-1mm}
\epsfig{figure=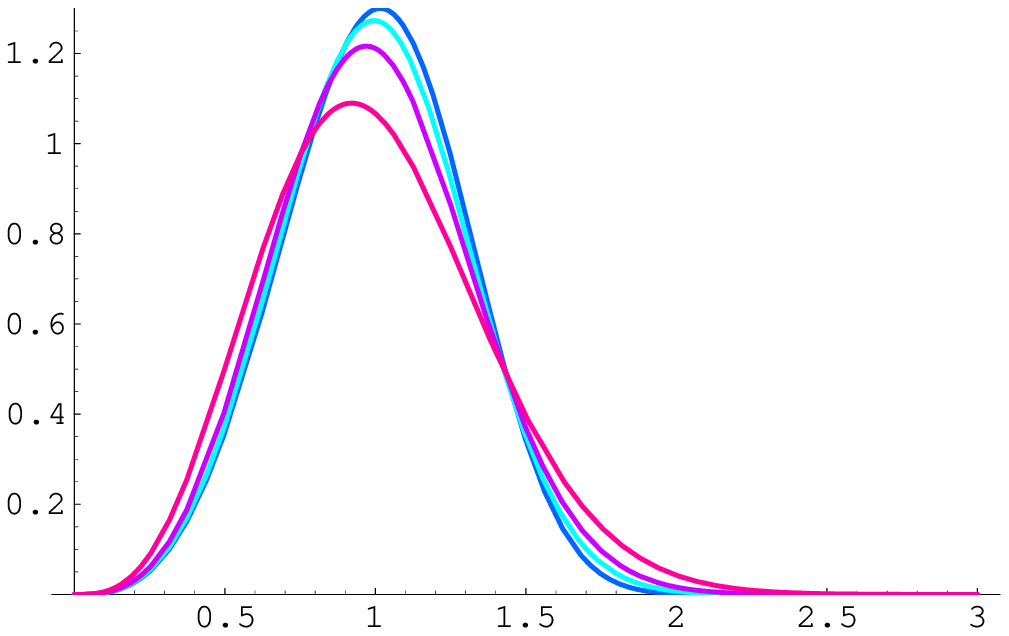,height=28mm}
 \put(-115,75){\tiny $p_{Gin}^{(2)}(r)$}
 \put(-135,10){\tiny $s$}
 \put(-245,75){\tiny $p_{W\!S}^{(\beta)}(s)$}
 \put(-5,10){\tiny $r$}
}
\caption{Left: surmise
$p_{W\!S}^{(\beta)}(s)$ 
(red) vs exact result (dashed blue) in \cite{DH}.
Right: $p_{Gin}^{(2)}(r)$ for $N=2,3,4,20$ (red to blue).
\\[-4ex]
}
\label{WSfig}
\end{figure}

The exact result $p^{(\beta)}(s)$ is
cumbersome, given in terms of an infinite product of eigenvalues of spheroidal
functions (e.g. in \cite{Mehta}), the 5th Painlev\'e transcendent \cite{Mehta},
or combining a Taylor series with coefficients given by sums over
permutations and Dyson's asymptotic expansion in a Pad\'e approximation
\cite{DH}. This is compared to the surmise Eq. (\ref{WS})
in Fig. \ref{WSfig} left.
In Table \ref{deltas}
we give the root of the integrated square deviation
for later comparison,
\be
\delta\equiv\Big[
\int_0^\infty ds\,(p^{(\beta)}(s)-p_{surmise}^{(\beta)}(s))^2\Big]^{\frac12}\ .
\label{deltadef}
\ee

The non-Hermitian WD ensembles are
defined by dropping the Hermiticity constraint in Eq. (\ref{ZWD}) left
\cite{Gin}.
We only display the complex eigenvalue representation
for $\beta=2$ and $4$ and their Jacobians ${\cal J}_\beta(z)$ computed in
\cite{Gin}:
\bea
{\cal Z}_{Gin}^{(\beta)}
\label{Jacobi}
&=&\int_{\mathbb{C}}\prod_{j=1}^N d^2z_j\ \mbox{e}^{-|z_j|^2}{\cal J}_\beta(z),
\\
{\cal J}_2(z)&=&|\Delta_N(z)|^2,\
{\cal  J}_4(z)=\Delta_{2N}(z,z^*)\prod_{j=1}^N(z_j-z_j^*).
\nn
\eea
For $\beta=2$ the spacing is obtained from an $N=2$ surmise by
inserting $\delta(|z_1-z_2|-r)$ in ${\cal Z}_{Gin}^{(2)}$,
and putting one eigenvalue at the origin.
The exact spacing for any $N$ obtained in \cite{GHS88}
uses translational invariance in the bulk
\be
\hat{p}_{Gin}^{(2)}(r)=-\frac{\partial E_{Gin}^{(2)}(r)}{\partial r}, \
E_{Gin}^{(2)}(r)=\! \prod_{j=1}^{N-1}\!\mbox{e}^{-r^2}\!
\sum_{k=0}^{j}\frac{r^{2k}}{k!}.
\label{pGin}
\ee
In Fig. \ref{WSfig} right we compare $N=2$ with increasing $N$, all
curves having norm and first moment 1. Clearly a surmise
{\it does not work} for the $\beta=2$
Ginibre ensemble ($\delta\approx 0.18$), as previously noted in \cite{GHS88}.
For $\beta=4$ and 1 the spacing is currently unknown.


\begin{table}[b]
\begin{tabular}{c||c||c||c|c|c||c|c|c}
    &$\delta_{W\!S}$ & &$\delta_{\mu=0}^{\nu=0}$ & $\delta_{\mu=0}^{\nu=1}$
& $\delta_{\mu=0}^{\nu=2}$  & $\delta_{\mu=1}^{\nu=0}$ &
  $\delta_{\mu=1}^{\nu=1}$ & $\delta_{\mu=1}^{\nu=2}$    \\
  \hline
GUE &0.04           & chGUE& 0       &$3.8$  & $7.7$& 
8.0 & 12.3 & 14.8 \\
GSE &0.015          & chGSE& $1.7$   & $6.1$ &  $10.6$ & 
1.8 & 3.3 & 4.4 \\
GOE &0.16           & chGOE& $3.6$   & 0     & -    & 
-& -&- \\
\end{tabular}
\caption{Deviation Eq. (\ref{deltadef}) in units $10^{-3}$
between approximate $N=2$ and exact large-$N$ results ($\delta_{W\!S}$ from
\cite{DH}).
}
\label{deltas}
\end{table}


3. {\it First eigenvalue in chiral RMT.}
The chiral ensembles with real eigenvalues called
chGOE, chGUE, and chGSE
are defined in terms of
$N\times(N+\nu)$ rectangular matrices $W$ with real, complex or quaternion real
elements without further symmetry restrictions. Switching to positive
eigenvalues $\la_j\geq0$ of the Hermitian Wishart (or covariance)
matrix $WW^\dag$ we obtain
\be
{\cal Z}_{\nu}^{(\beta)}\!=\!
\int_0^\infty\!\prod_{j=1}^N d\la_j\la_j^{d}\mbox{e}^{-\la_j}
|\Delta_N(\la)|^\beta,\ \mbox{$d\equiv \frac{\beta(\nu+1)}{2}-1$}.
\label{Zch}
\ee
Here $N_f$ massless flavours can be added by shifting $d\to d+N_f$.
The gap probability $E^{(\beta)}(s)$
that the interval $(0,s)$ is void follows by integrating in Eq. (\ref{Zch})
from $s$ to $\infty$. For $N=2$  we obtain
\be
E_\nu^{(\beta)}(s)\sim\!\! \int_0^\infty\!\!\!\!
dxdy[(x+s)(y+x+s)]^{d}\mbox{e}^{-2(s+x)-y} y^\beta,
\label{EN2}
\ee
after shifting variables.
The nested integrals can easily be evaluated.
Note that $d\!=\!0$ for $\beta\!=\!2$, $\nu\!=0$, and $\beta\!=\!1\!=\!\nu$.
These gap probabilities can be computed exactly for any $N$,
and our surmise gives the {\it exact} result
after rescaling.

To compare with Dirac operator eigenvalues
we have to switch variables $\la_j\to y_j^2$,
coming in eigenvalue pairs
$\pm y_j$, and thus to $s\to s^2$.
The distribution of the first positive Dirac eigenvalue follows:
$p_{\nu}^{(\beta)}(s)
=-\partial_s [E_\nu^{(\beta)}(s^2)]$.

We first list all its known $N_f\!=\!0$ results in the universal 
microscopic limit
for $\nu\in\mathbb{N}$ in Eqs. (\ref{pb2}) - (\ref{pb4}): the
chGUE for all $\nu$ \cite{WGW,DN98},
the chGOE for $\nu=0$ \cite{Forrester1st} and odd $\nu$ \cite{DN98},
and the chGSE for $\nu=0$ \cite{Forrester1st} and $\nu>0$ \cite{BBMW-1998}.
For the latter, only a convergent
Taylor series is known with coefficients $a_j(\nu)$ given by
sums over partitions (see Eq. (8) in \cite{BBMW-1998}),
much alike for the WS in the WD class,
\bea
p_{\nu}^{(2)}(s)&=& \,s\ \mbox{e}^{-s^2/4}\det_{i,j=1,\ldots,\nu}
\left[I_{i-j+2}(s)\right]/2,
\label{pb2}\\
{p}_{\nu=0}^{(1)}(s)&=& [(2 + s)\,\mbox{e}^{-s^2/8-s/2}]/4,
\label{pb1}\\
\hat{p}_{\nu=2n+1}^{(1)}(s)&\sim&s^{(3-\nu)/2}\mbox{e}^{-s^2/8}
\stackrel{\Pf\ \ \left[ (i-j)I_{i+j+3}(s)\right]}{
\mbox{\tiny$i,j=-n+\frac12,\ldots, n-\frac12$}}
\ \ ,
\nn\\
{p}_{\nu=0}^{(4)}(s)&=&(\pi/2)^{\frac12}s^{\frac32}\mbox{e}^{-s^2/2}
I_{3/2}(s), \label{pb4}\\
\hat{p}_{\nu>0}^{(4)}(s)&\sim&
s^{4\nu+3} \mbox{e}^{-s^2/2}
\Big( 1+ \sum_{j=1}^\infty{a_j(\nu)s^j}\Big).
\nn
\eea
Next, we give examples following our surmise Eq. (\ref{EN2})  where
$p_{\nu}^{\beta}(s)$ is {\it not} known in elementary form, filling the
gaps in Eqs. (\ref{pb2}) - (\ref{pb4}) for the first two values of $\nu>0$:
\bea
\hat{p}_{\nu=2}^{(1)}(s)&\sim&{3s^3}\mbox{e}^{-\frac{s^2}{8}}
+ \Big(6s^2 - \frac{s^4}{4}\Big)
\mbox{e}^{-\frac{1}{16}s^2}\! \sqrt{\pi} \mbox{Erfc}\Big[\frac{s}{4}\Big],
\label{pb1nu2N2}
\\
\hat{p}_{\nu=4}^{(1)}(s)&\sim&\!\!
(s^5 +\frac{s^7}{60})\mbox{e}^{-\frac18 s^2}\!\!\!+\!
(2s^4 - \frac{s^6}{20})
     \mbox{e}^{-\frac{1}{16}s^2}
\!\!\sqrt{\pi} \mbox{Erfc}\Big[\frac{s}{4}\Big],
\nn
\\
\hat{p}_{\nu=1}^{(4)}(s)&\sim&
s^7 (13440 + 1440 s^2 + 60s^4 + s^6)
\mbox{e}^{-\frac12s^2}, 
\label{pb4nu1N2}
\\
\hat{p}_{\nu=2}^{(4)}(s)&\sim&
s^{11}
(15482880 + 2150400 s^2 + 134400 s^4 \nn\\
&&+ 4800 s^6 + 100 s^8 +   s^{10})
\mbox{e}^{-\frac12s^2} 
.
\nn
\eea
The normalisation constants suppressed above easily follow.
However, we cannot  set the 1st
moment to one as in the WD class. The position of $p_{\nu}^{(\beta)}(s)$
measures the repulsion by $\nu$ exact zero-eigenvalues, containing
important information.
Thus we fix the $N=2$ scale by setting the 1st moment equal to the exact one.
Without exact ($\beta=1$, even $\nu$) or concise ($\beta=4$, $\nu>0$)
results we instead fit to
the increasing slope of the known microscopic density
$\rho_\nu^{(\beta)}(s)$, being the first term in the
Fredholm expansion of the 1st eigenvalue \cite{AD03} (see also
Eq. (\ref{Fredexp2})).
\begin{figure}[t]
\centerline{
  \epsfig{figure=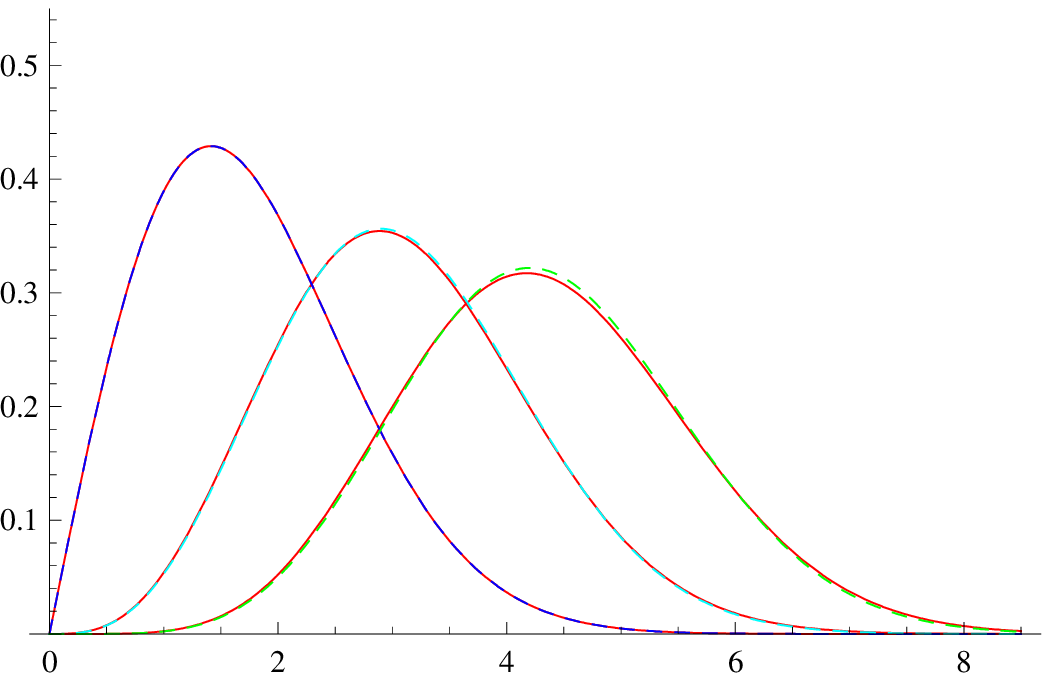,height=28mm}\hspace*{-1mm}
  \epsfig{figure=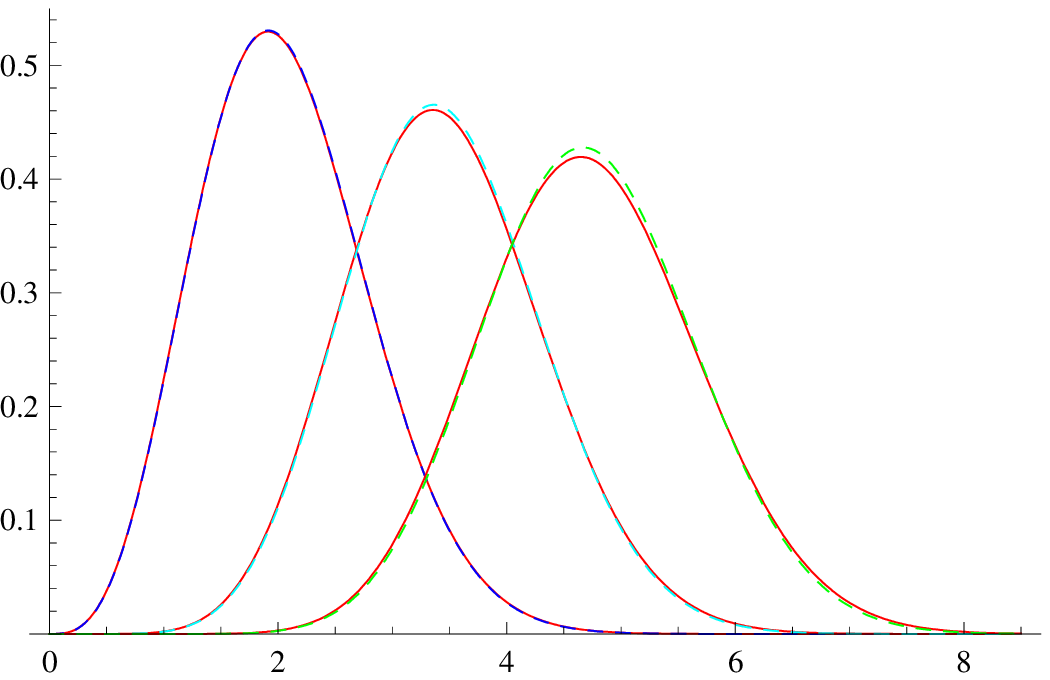,height=28mm}
 \put(-115,79){\tiny $p_{\nu}^{(4)}(s)$}
 \put(-135,10){\tiny $s$}
 \put(-238,79){\tiny $p_{\nu}^{(2)}(s)$}
 \put(-5,10){\tiny $s$}
}
\centerline{\epsfig{figure=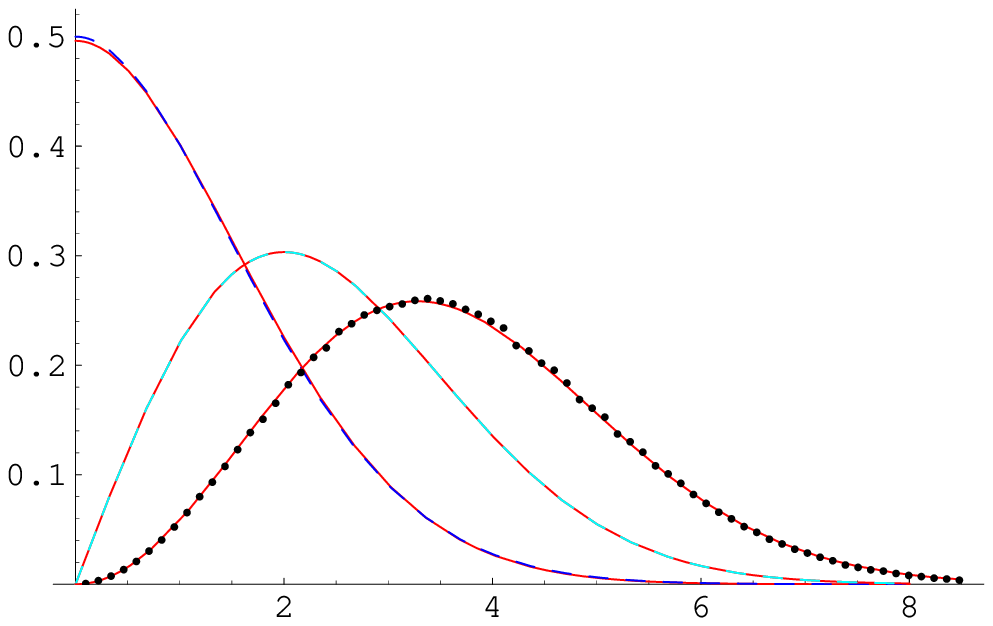,height=28mm}\hspace*{-1mm}
\put(-5,10){\tiny $s$}
\put(-105,75){\tiny $p_{\nu}^{(1)}(s)$}
}
\caption{$p_\nu^{(\beta)}(s)$ with
$\nu=0,1,2$ in dashed blue to green for $\beta=2$ (top left), $\beta=4$
(top right) and $\beta=1$ (bottom). The $N=2$ surmise is in red. Our new
result for $\nu=2$ at $\beta=1$ is compared to a numerical simulation at
$N=20$ (black dots).
\\[-4ex]
}
\label{pchplot}
\end{figure}
In Fig. \ref{pchplot} we compare
approximate to exact 1st eigenvalues for small topology $\nu=0,1,2$ and all
$\beta$.
The deviation measured by
Eq. (\ref{deltadef}) in Table \ref{deltas}
increases with $\nu$, becoming
visible only for $\nu=2$ (see Fig. \ref{pchplot}).
This has to be compared to the statistical error in data, see e.g. Fig.
\ref{3Dplot}.

Note that in chiral RMT the nn spacing also obeys Eq. (\ref{WS}),
but {\it does not follow} from an $N\!=\!2$ 
surmise \cite{AAV}.

\begin{figure}[h]
\centerline{
  \epsfig{figure=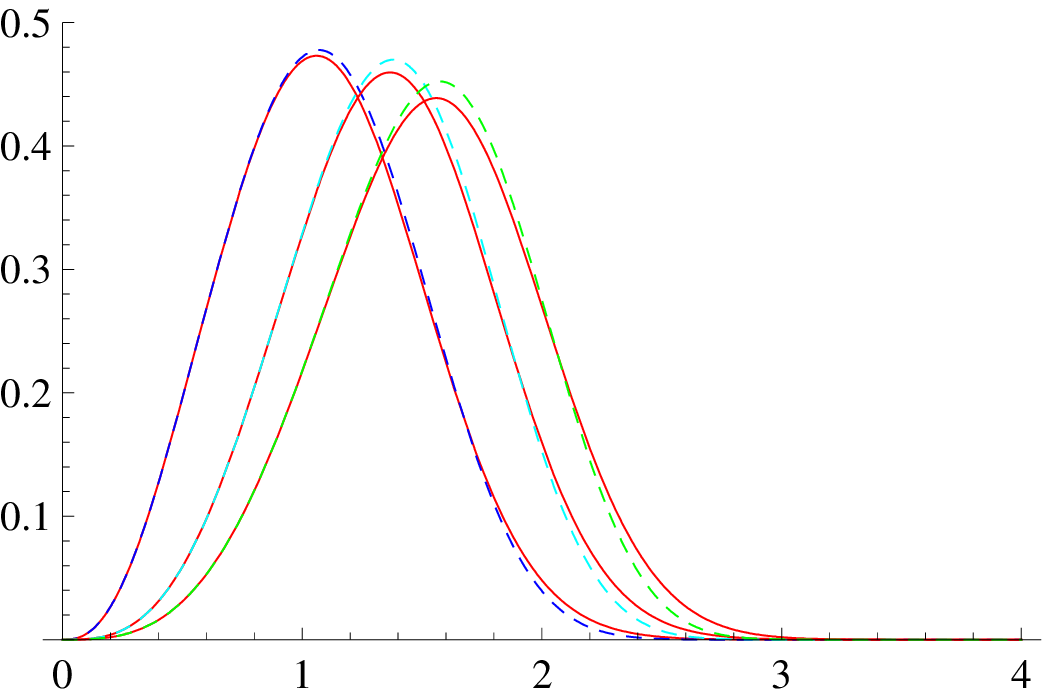,height=28mm}
\hspace*{-1mm}
  \epsfig{figure=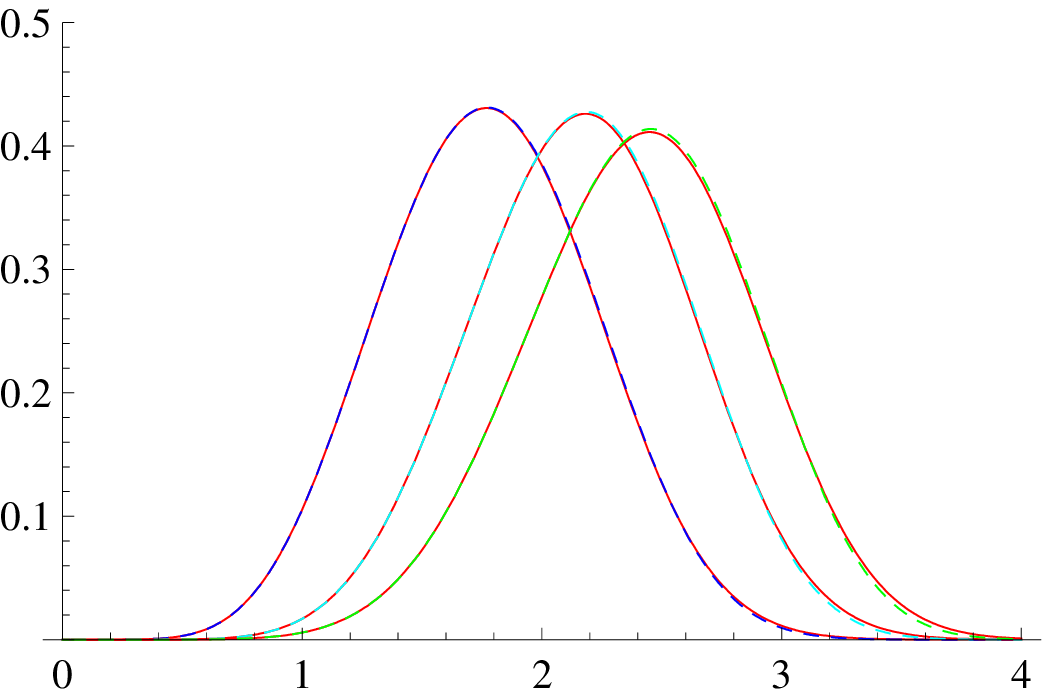,height=28mm}
 \put(-108,80){\tiny $\int\!\! d\theta p_\nu^{(4)}(re^{i\theta})$}
 \put(-130,10){\tiny $r$}
 \put(-230,80){\tiny $\int\!\! d\theta p_\nu^{(2)}(re^{i\theta})$}
 \put(-5,10){\tiny $r$}
}
\caption{Integrated 1st eigenvalue at $\mu\!=\!1$
for $\nu\!=\!0,1,2$ (blue to green dashes) vs $N\!=\!2$ (red): 
$\beta\!=\!2$ (left) and $\beta\!=\!4$ (right).
\\[-4ex]
}
\label{pchGinplot}
\end{figure}

The non-Hermitian  chiral
ensembles with $\mu\neq0$
are given in terms of a two-matrix model \cite{James,A05}.
We only focus on $\beta=2,4$ here,
with their complex eigenvalue representations for
$N_f=0$ reading
\cite{James,A05}
\be
{\cal Z}_{\nu\,\mathbb{C}}^{(\beta)}\!=\!\!
\int_{\mathbb{C}}\prod_{j=1}^N d^2z_j
|z_j|^{\beta\nu+2}K_{\frac{\beta\nu}{2}}(a|z_j|^2)\mbox{e}^{b\re z_j^2}
{\cal J}_\beta(z^2).\!
\label{ZchG}
\ee
The weight $w(z)$ depends on $a\equiv\frac{1+\mu^2}{2\mu^2}>b\equiv
\frac{1-\mu^2}{2\mu^2}\geq0$, with $\mu\in(0,1]$. The limit $\mu\to0$ leads
back to real eigenvalues, and at $\mu=1$  non-Hermiticity is maximal.
The definition of a gap probability on $\mathbb{C}$ is not unique
\cite{ABSW,APS}. For \textit{radial} ordering
it reads
\be
E^{(\beta)}(r)\sim
\prod_{j=1}^N \int_r^\infty dr_jr_j\int_0^{2\pi} d\theta_j
w(z_j){\cal J}_\beta(z)\ .
\label{gapdef}
\ee
Differentiation yields $\partial_rE^{(\beta)}(r)=
\int_0^{2\pi}d\theta \, p_\nu^{(\beta)}(re^{i\theta})$, the integrated 1st
eigenvalue. For $\beta=2$ (4) the gap probability is
given by a Fredholm determinant (Pfaffian) \cite{APS}
\be
E^{(2)}(r) \sim\! \det_{1,\ldots,N}
\Big[\int_{r^2}^\infty\!\!\! dt\,t^{k+j+\nu-1} K_\nu(at) I_{k+j-2}(bt)\Big]\!.
\!
\label{fred}
\ee
Its
matrix elements $A_{jk}^{(\nu)}$ can be computed recursively for any $\nu$
by differenting the following matrix element \cite{APS}:
\be
A_{11}^{(0)}
=\frac{br^2I_{1}(br^2) K_0(ar^2)+ar^2I_{0}(br^2) K_1(ar^2)}{a^2-b^2} \ .
\ee
This leads to a
$\beta\times\beta$ determinant (Pfaffian) representation for our $N=2$ surmise
valid for any $\mu$. At $\mu=1$ all Fredholm eigenvalues
$1-\la_{k=0,\ldots,N-1}^{(\beta)}$ are
explicitly known \cite{APS}, providing an exact result for any $N$ as in
Eq. (\ref{pGin}). It contains incomplete Bessel function series
$I_\nu^{[k]}(x)$ truncated at power $k$ ($\equiv0$ for $k<0$)
\bea
&&\!\!\!\!\!\!\!\!(1-\la_k^{(2)})
=
\frac{r^{2(2k+\nu+1)}}{2^{2k+\nu}(k+\nu)!k!}K_{\nu+1}(r^2)
\label{frEig}\\
&&+r^2(I_{\nu+2}^{[k-2]}(r^2)K_{\nu+1}(r^2)
+I_{\nu+1}^{[k-1]}(r^2)K_{\nu+2}(r^2)).
\nn
\eea
For $\beta=4$ we have
the relation $\la_k^{(4)}=\la_{2k+1}^{(2)}$ with $\nu\to2\nu$ \cite{APS}.
In Fig. \ref{pchGinplot} we compare our surmise to this result, truncated
at $N\!=\!8$ because of rapid convergence. Here it works better for
$\beta\!=\!4$ than $\beta\!=\!2$, in contrast to $\mu\!=\!0$. 
Due to angular integration only one
scale has to be fixed after
normalisation, which can be done as in the real case.

Next we give a surmise for $p_\nu^{(\beta)}(re^{i\theta})$.
In Eq. (\ref{gapdef})
we skip the integration over $\theta_1$ and differentiate wrt $r_1$.
For $N=2$  we obtain
an exact Fredholm expansion
\be
p_{\nu}^{(\beta)}(z)= R_{1,\nu}^{(\beta)}(z)-\int_0^{r_1} dt \, t
\int_0^{2\pi} d\varphi
R_{2,\nu}^{(\beta)}(z,t\mbox{e}^{i\varphi})\ ,
\label{Fredexp2}
\ee
with $z\!=\!r_1\mbox{e}^{i\theta_1}\!=\!x+iy$.
The 1- and 2-point spectral densities are expressed through the
kernel of orthogonal Laguerre polynomials of norm $h_j$ (see \cite{James} for
details)
\be
R_{1,\nu}^{(2)}(z)=K_N^{(2)}(z,z^*)= w(z)\sum_{j=0}^{N-1}
\frac{{\Big|L_j^{(\nu)}\Big(\frac{z^2}{1-\mu^2}\Big)\Big|^2}}{h_j},
\label{densities}
\ee
and $R_{2,\nu}^{(2)}(z,u)=R_{1,\nu}^{(2)}(z)
R_{1,\nu}^{(2)}(u)-|K_N^{(2)}(z,u^*)|^2$.
For $\beta\!=\!4$ we have a Pfaffian of a matrix kernel  instead \cite{A05}.
An example for $p_{\nu=0}^{(4)}(z)$ is shown in  Fig.\,\ref{3Dplot} top right.
Here two scales have to be fixed: 
for $z$ we fit to the increase of the known microscopic density
in the $x$-direction, and for rescaling
$2N\mu^2\!\equiv\! \alpha^2$ to its decrease in the $y$-direction.
Since $\alpha\!\leq\!2$ 
for $N\!=\!2$, we conclude that at large-$N$ for $\alpha\!>\!2$ 
$p_{\nu}^{(\beta)}(z)$
must become symmetric wrt rotation ($\beta\!=\!2$) or reflections
wrt the bisector of each quadrant ($\beta\!=\!4$). We have checked this, 
as well as distributions 
for $0\!<\!\mu\!<\!1$ by generating
ensembles of large random matrices.
\begin{figure}[t]
\centerline{
  \epsfig{figure=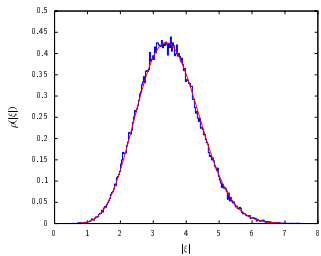,height=33mm}
}
\caption{$\int\!\! d\theta \, p_\nu^{(4)}(re^{i\theta})$ (red)
vs Lattice data \cite{AB} (blue) with
volume $V=4^4$, gauge coupling  1.3,
$\mu_{\rm Lat}=0.2$ and mass $20$ in Lattice units, using
a very large number $10^5$ configurations.
\\[-4ex]
}
\label{2Dplot}
\end{figure}

4. {\it Lattice data.}
In \cite{AB} two-colour QCD was
compared to the $\beta=4$
microscopic spectral density in the complex plane from chiral RMT \cite{A05}.
We use the same data here but with higher statistics, and refer to \cite{AB}
for all simulation details. Because unimproved staggered
fermions are used we are in the $\beta\!=\!4$ class at $\nu\!=\!0$. Our 
$N_f\!=\!2$
data are effectively quenched for the smallest eigenvalues due to a
large mass.  In Fig. \ref{2Dplot} we compare to the 1st integrated eigenvalue,
with $\alpha\!=\!1.352$ being close to maximal non-Hermiticity. No further fits
compared to \cite{AB} are made.
\begin{figure}[bp!]
\centerline{
  \epsfig{figure=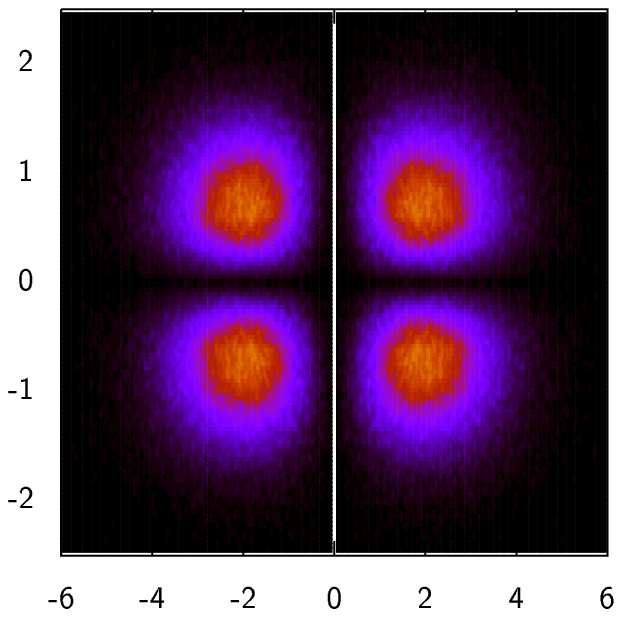,height=30mm}
  \epsfig{figure=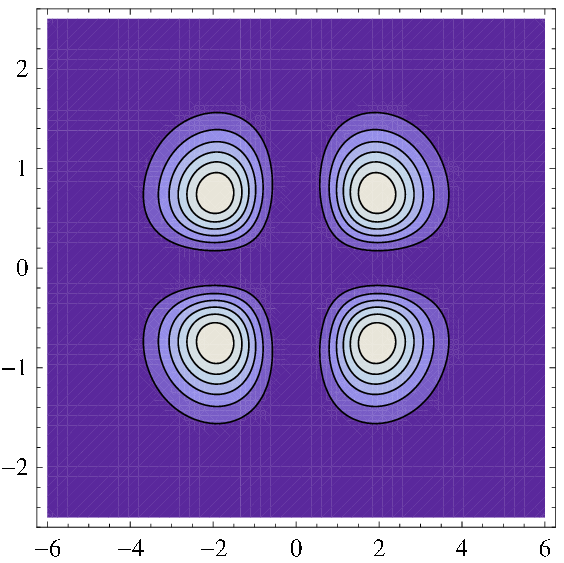,height=30mm}
}
\centerline{
  \epsfig{figure=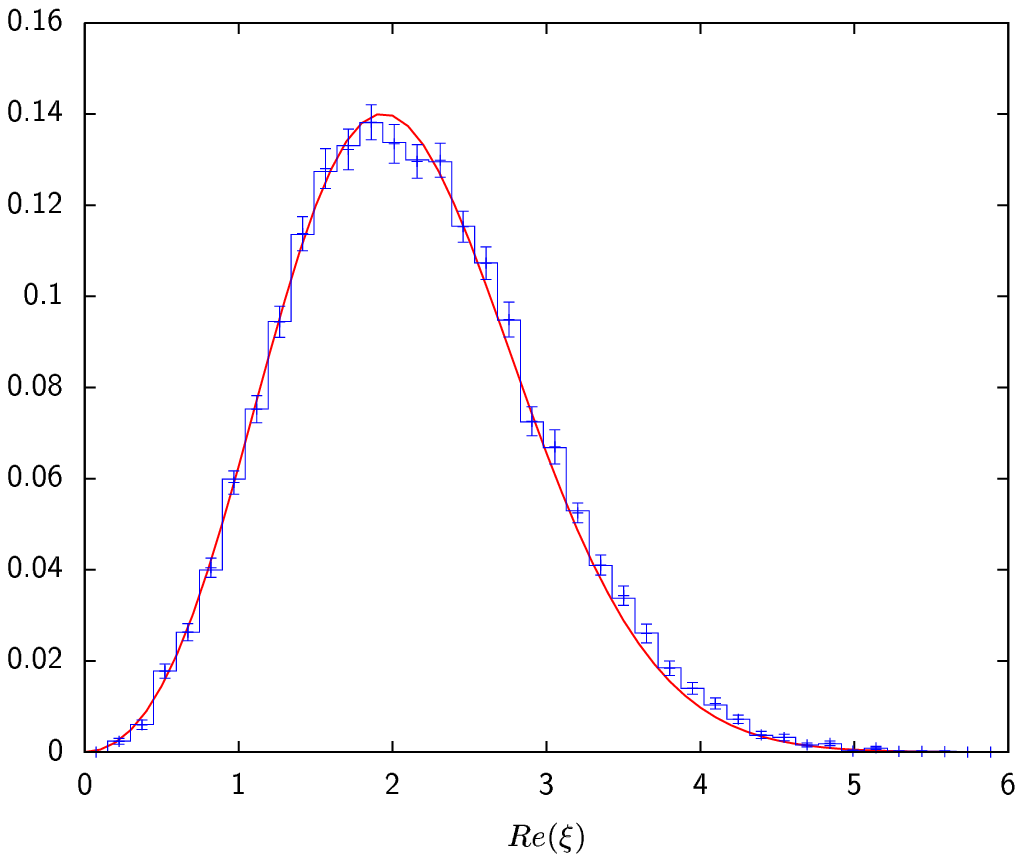,height=30mm}\hspace*{-1mm}
  \epsfig{figure=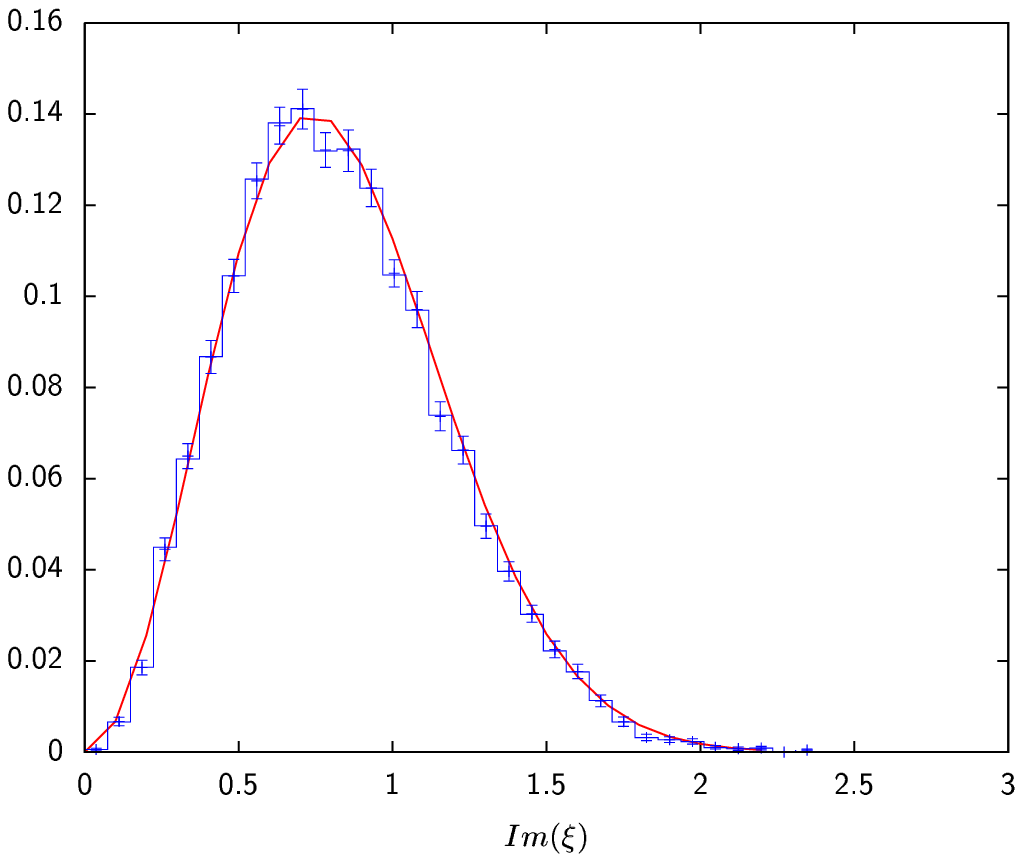,height=30mm}
}
\caption{Top: contour plots for Lattice data as in Fig. \ref{2Dplot} but with
  $\mu_{\rm Lat}=0.1$ (left) vs surmise Eq. (\ref{Fredexp2}) (right).
Bottom: cuts through a single peak in $x$- (left) and $y$-direction (left).
\\[-4ex]
}
\label{3Dplot}
\end{figure}
In Fig. \ref{3Dplot} we compare LGT data at intermediate $\mu_{\rm Lat}=0.1$
to the angle-dependent surmise Eq. (\ref{Fredexp2}) by taking cuts.
Here the two scales are fitted to the
data, finding an excellent agreement for $\alpha\!=\!0.65$.

An alternative to Eq. (\ref{Fredexp2}) is the
truncated Fredholm expansion in the microscopic large-$N$ limit \cite{AD03}
which was successfully applied to the $\beta\!=\!2$ class \cite{ABSW}.
However,  integrals of higher order terms rapidly become cumbersome.

5. {\it Conclusions.}
Conceptually it is possible within
chiral RMT to approximate the 1st eigenvalue distribution using a
$2\times(2+\nu)$ matrix calculation,
for both real and complex eigenvalues. It is remarkable that
this surmise works and
captures the repulsion of $\nu$ zero-eigenvalues. We derived
new compact expressions for $\beta=1$ and $4$ with
real eigenvalues for $\nu>0$.
Second, we have shown that our surmise for  $\beta=4$
successfully describes 
$SU(2)$ Lattice data, in an intermediate regime for $\mu\neq0$ where no
results were previously known.
It would be very interesting to extend our results to the $\beta=1$
non-Hermitian chiral class, having both real and complex eigenvalues.

Support by 
ENRAGE MRTN-CT-2004-005616 (G.A., E.B.), EPSRC grant
EP/D031613/1 (G.A., L.S.) and DFG grant JA483/22-1 (E.B.)
is acknowledged.
\\[-6ex]


\begin{thebibliography}{99}
\bibitem{GMW}T. Guhr, A. M\"{u}ller-Groeling, H.A.
Weidenm\"{u}ller, Phys. Rep. {\bf 299} (1998) 190.
\bibitem{Mehta}
M.L. Mehta, {\it Random Matrices}, Academic Press, Third
Edition, London 2004.

\bibitem{Gin} J. Ginibre, J. Math. Phys. {\bf 6} (1965) 440.

\bibitem{reviewFS} Y.V.\,Fyodorov,\,H.J.\,Sommers, J.\,Phys.\,{\bf
  A36}\,(2003)\,3303.

\bibitem{MPW}
  H. Markum, R. Pullirsch, T. Wettig,
  Phys. Rev. Lett.  {\bf 83} (1999) 484.

\bibitem{SV93}
E.V. Shuryak, J.J.M.\,Verbaarschot, Nucl. Phys. {\bf A560} (1993) 306.

\bibitem{Steph} M. Stephanov, { Phys. Rev. Lett.} {\bf 76} (1996) 4472.

\bibitem{Edwards}
R.G. Edwards, U.M. Heller, J.E. Kiskis, R. Narayanan,
Phys. Rev. Lett. {\bf 82} (1999) 4188.

\bibitem{Bloch}
  J. Bloch, T. Wettig,
Phys. Rev. Lett. {\bf 97} (2006) 012003.

\bibitem{QCDmu} S. Ejiri, PoS (LATTICE 2008) 002.

\bibitem{James} J.C. Osborn,
Phys. Rev. Lett. {\bf 93} (2004) 222001.

\bibitem{AOSV}
G. Akemann, J.C. Osborn, K. Splittorff, J.J.M. Verbaarschot
Nucl. Phys. {\bf B712} (2005) 287.

\bibitem{KimJac} K. Splittorff, J.J.M. Verbaarschot,
Phys. Rev. Lett. {\bf 98} (2007) 031601;
Phys. Rev. {\bf D75} (2007) 116003.

\bibitem{ABSW}
G. Akemann, J. Bloch, L. Shifrin, T. Wettig,
Phys. Rev. Lett. {\bf 100} (2008) 032002.

\bibitem{DN98}
P.H. Damgaard, S.M. Nishigaki,
  Phys. Rev. {\bf D63} (2001) 045012.

\bibitem{ITEP}
P.V. Buividovich, E.V. Luschevskaya, M.I. Polikarpov,
Phys. Rev. {\bf D78} (2008) 074505.

\bibitem{AB}
G. Akemann, E. Bittner,
  Phys. Rev. Lett. {\bf 96} (2006) 222002.

\bibitem{AD03}
  G. Akemann, P.H. Damgaard, Phys. Lett. {\bf B583} (2004) 199.

\bibitem{APS} G. Akemann, M.J. Phillips, L. Shifrin,
J. Math. Phys. {\bf 50}  (2009) 063504.

\bibitem{DH} B. Dietz, F. Haake, Z. Phys. {\bf B80} 
  (1990) 153. 

\bibitem{GHS88} R. Grobe, F. Haake, H.J. Sommers,
  Phys. Rev. Lett. {\bf 61} (1988) 1899.

\bibitem{WGW}
T. Wilke, T. Guhr, T. Wettig,
  Phys.  Rev. {\bf D57} (1998) 6486;
S.M. Nishigaki, P.H. Damgaard, T. Wettig,
Phys. Rev. {\bf D58} (1998) 087704.


\bibitem{Forrester1st} P.J. Forrester, Nucl. Phys. {\bf B402} 
(1993) 709.


\bibitem{BBMW-1998} M.E. Berbenni-Bitsch, S. Meyer, T. Wettig,
Phys. Rev. {\bf D58} (1998) 071502.

\bibitem{AAV} A.Y. Abul-Magd, G. Akemann, P. Vivo,
J. Phys. {\bf A42}
(2009) 175207.

\bibitem{A05}
G. Akemann,
  Nucl. Phys. {\bf B730} (2005) 253.


\end{thebibliography}
\end{document}